\documentclass[iop]{emulateapj}
\usepackage{txfonts}
\usepackage{graphicx}
\usepackage{natbib}

\def\etal{{\sl et al.}}
\usepackage{natbib}
\shorttitle{INVESTIGATING IRIS/ELLERMAN BOMBS IN 1-D RADIATIVE HYDRODYNAMICS}
\shortauthors{Reid et al.}

\begin{document}

\title{SOLAR ELLERMAN BOMBS IN 1-D RADIATIVE HYDRODYNAMICS}
\vskip1.0truecm
\author{
A. Reid$^{1,4}$, M. Mathioudakis$^{1}$, A. Kowalski${^{2, 3}}$, J. G. Doyle${^4}$, J. C. Allred${^5}$}
\affil{
1. Astrophysics Research Centre, School of Mathematics and Physics, Queen's University Belfast, BT7~1NN, Northern Ireland, UK; e-mail: aaron.reid@qub.ac.uk\\
2. Department of Astrophysical and Planetary Sciences, University of Colorado Boulder, 2000 Colorado Ave, Boulder, CO 80305, USA. \\
3. National Solar Observatory, University of Colorado Boulder, 3665 Discovery Drive, Boulder, CO 80303, USA \\
4. Armagh Observatory and Planetarium, College Hill, Armagh, BT61 9DG, UK\\
5. NASA/Goddard Space Flight Center, Code 671, Greenbelt, MD 20771\\
}
%

\begin{abstract}
Recent observations from the Interface Region Imaging Spectrograph (IRIS) appear to show impulsive brightenings in high temperature lines, which when combined with simultaneous ground based observations in H$\alpha$, appear co-spatial to Ellerman Bombs (EBs). We use the RADYN 1-dimensional radiative transfer code in an attempt to try and reproduce the observed line profiles and simulate the atmospheric conditions of these events. Combined with the MULTI/RH line synthesis codes, we compute the H$\alpha$, Ca II 8542~\AA, and Mg II h \& k lines for these simulated events and compare them to previous observations. Our findings hint that the presence of superheated regions in the photosphere ($>$10,000 K) is not a plausible explanation for the production of EB signatures. While we are able to recreate EB-like line profiles in H$\alpha$, Ca II 8542~\AA, and Mg II h \& k, we cannot achieve agreement with all of these simultaneously. 

\end{abstract}


\section{INTRODUCTION}

\begin{figure*}[!t]
\plotone{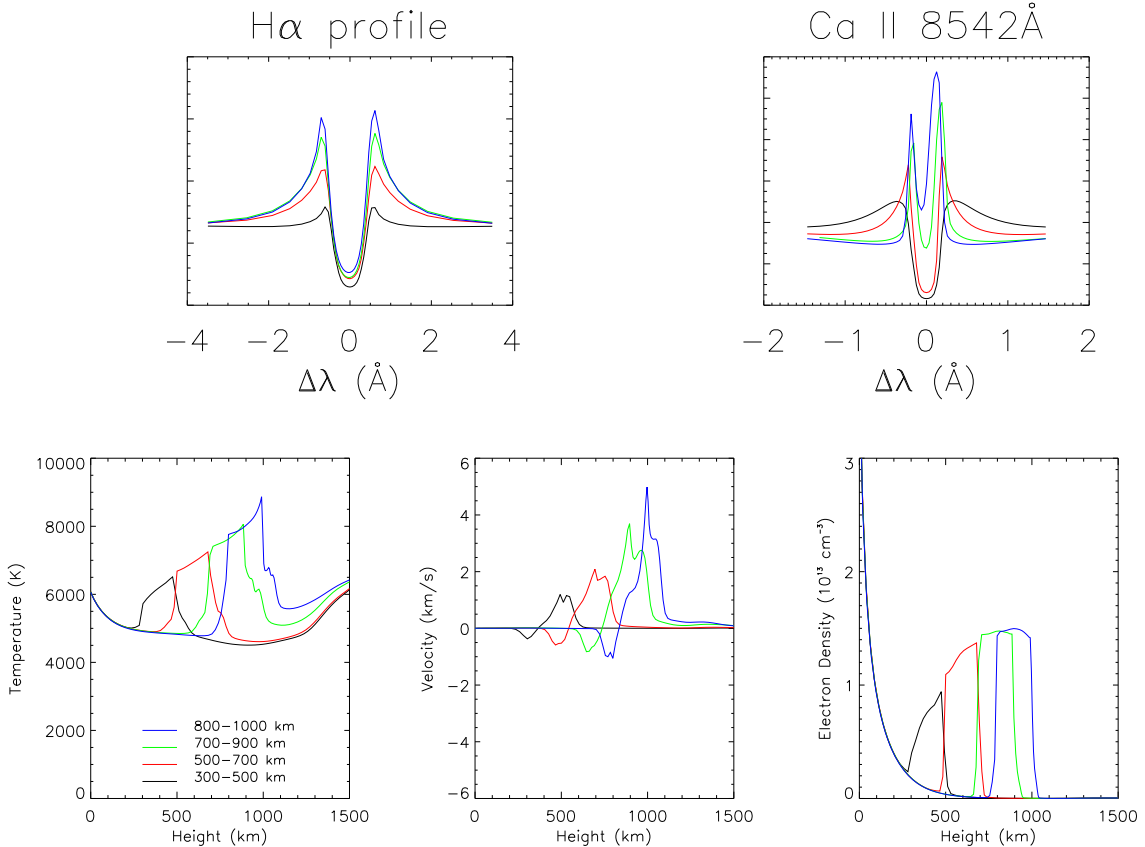}

\caption{Top Left: Synthesised H$\alpha$ line profiles for energies deposited at various heights, using the quiet Sun starting atmosphere. Top Right: The corresponding Ca II 8542~\AA\ line profiles. Bottom Left: Temperature profiles across the temperature minimum region. Bottom Center: Vertical velocity profiles. Bottom Right: Electron density profiles.}
\label{fig1}
\end{figure*}
Ellerman Bombs (EBs henceforth) were first noticed in \cite{ell}, who noted brightenings in the wings of the H$\alpha$, H$\beta$, and H$\gamma$ lines. These events have also been observed in the wings of Ca II 8542~\AA\ \citep{fang, socas4, par2, viss2, li}, as well as in Ca II H images \citep{matsu2, hashi}. They show no observable enhancement in the core of these line profiles as these are formed in the overlying chromospheric canopy. EBs are considered a solely photospheric/lower chromospheric phenomenon \citep{viss}.

EB brightenings are also observable in the Solar Dynamics Observatory (SDO) 1700~\AA\ and 1600~\AA\ channels \citep{qiu, geo, par2, berlicki2, viss}, though to a lesser degree than the H$\alpha$ line wings due to the broad passbands of these filters encompassing a wide range of atmospheric heights \citep{viss}. While the 1600~\AA\ channel offers better contrast than the 1700~\AA\ channel \citep{rutt, viss}, EB signatures are more difficult to observe in the 1600~\AA\ channel due to contamination effects from C IV emission with transition region temperatures.

EBs typically last for a few minutes, and appear rather impulsively in comparison with other photospheric brightenings such as moving magnetic features (MMFs). \cite{reid2016} have shown that by introducing an impulsivity criterion, it is possible to distinguish between pseudo-EBs and EBs. EBs are generally observed with co-spatial blue-shifts, or bi-directional Doppler shifts \citep{matsu2, wat1}.

EBs are generally found in regions of opposite polarity magnetic flux \citep{geo, wat, matsu, hashi, Nelson1, viss}. Recent spectropolarimetric inversions of these events show evidence for flux cancellation \citep{reid2016}, with magnetic energies comparable to that of the radiative energy losses. EBs have been estimated to form in the temperature minimum region \citep{Nelson3}, with foot-points reported to form as low as 300~km \citep{wat1}. As such, they are thought to appear due to photospheric magnetic reconnection, occurring around the temperature minimum region, where this process is most efficient \citep{lit2}. 

Three-dimensional numerical modeling of photospheric reconnection has shown local temperature increases in the photosphere by a factor of 1.1-1.5 relative to quiet Sun, along with a density increase by a factor of 4 at the magnetic inversion line \citep{ach}. The \cite{ach} model has also shown bi-directional flows in the region, with values of 2-4 km s$^{-1}$.  Semi-empirical models for EBs show localized temperature enhancements of 600-3000~K around the temperature minimum region \citep{fang, berlicki}. These temperature enhancements lead to intensity enhancements in the wings of the H$\alpha$ and Ca II 8542~\AA\ lines.  Other EB studies also find similar temperature enhancements (200-3000K) in the photosphere/temperature minimum region \citep{geo, iso, yang,hong, li}.

EBs have also been found within quiet Sun regions of the photosphere \citep{luc}. The observational signatures of these QSEBs are not as pronounced as those found in the vicinity of active regions, though they are still located in regions of opposite polarity magnetic flux. This suggests that these events are identical to EBs occurring at locations of weaker magnetic flux cancellation. 

 More recently, observations using the IRIS instrument indicate that EB signatures have been found in the Mg II h \& k wings, Si IV, and C II lines \citep{Peter, kim, viss2, tian}. These lines are sensitive to much higher temperatures (50,000 - 100,000 K), and sample the upper chromosphere and transition region. \cite{viss2} suggest that they are formed below the chromospheric canopy due to superheating with temperatures up to 80,000~K. \cite{judge} has debated the origins of these ‘bombs’ on the basis that the UV photons detected in the Si IV line cannot escape if formed below 500~km above the photospheric floor.

The MURaM code used by \cite{reid} shows H$\alpha$ wing enhancement at the magnetic inversion line of a bipolar structure, co-spatial with temperature enhancements. This event has shown flux cancellation, though the simulation was of quiet Sun, and so was more likely a QSEB. With the strong variance in temperatures required to produce the EB signatures, synthesising these line profiles from hydrodynamical simulations can help clarify the processes involved and also offers an alternative approach to EB line formation modeling.   

In this study, we use the RADYN code to model an EB-like atmosphere, with corresponding synthesis of the H$\alpha$ (Section 2), Ca II 8542~\AA\ (Section 2), and Mg II h \& k (Section 3) line profiles for comparison with previous observational studies (e.g. \cite{viss2}). 

\section{H$\alpha$ and Ca II 8542~\AA\ modeling}
The RADYN code \citep{carlsson1992, carlsson1994, carlsson1995} has been used extensively to study flare dynamics via beam heating in a 1-dimensional solar atmosphere. \cite{rubio} used the RADYN code in an attempt to model the atmosphere of an X1.0 solar flare, synthesising the H$\alpha$, Ca II 8542~\AA, and Mg II h \& k line profiles to compare with observations. The code also has the ability to apply a time dependent heating function into the atmosphere, and thus is aptly suited for the study of EBs. 

Our RADYN simulations use the quiet Sun starting atmosphere (QS.SL.LT in \cite{allred}). A time dependent heating function is applied to the atmosphere, allowing for a range of energy deposition rates over various portions of the photosphere/lower chromosphere. The MULTI line synthesis code is used to synthesise the H$\alpha$ and Ca II 8542~\AA\ lines to attempt to replicate the signatures of the observed line profiles with emission in the wings, while leaving the line cores unchanged. 

Previous estimates of EB energies are in the range of 10$^{24}$ - 10$^{27}$ ergs \citep{geo, fang, li, reid2016}. If we assume a lifetime of $\sim$5 minutes, and an active reconnection area of 300~km$^{2}$, the energy deposition rate would need to be of the order of 100 - 1000 ergs/cm$^{3}$/sec, assuming a vertical extent of $\sim$200~km. 

A grid of models was set up, applying heating rates between these values over 200~km in the photosphere (see Grid \textbf{1} from Table \ref{table1}). The atmosphere takes roughly 9 seconds to stabilize after the heating is applied, with all measurements taken at T=10s. A temperature enhancement appears at the location of energy deposition, ranging between 300 - 2600~K, with 2600~K relating to the 1000 ergs/cm$^{3}$/sec deposition rate. This is accompanied by an associated local electron density enhancement. Due to the sudden injection of energy, a shock forms at the deposition location. This creates bi-directional velocity flows up to 2 km s$^{-1}$. The upward velocity is at the leading edge of the shock and is stronger than the weaker, trailing downflow.

 With the associated synthesised H$\alpha$ line profiles, it was apparent that although the 100 ergs/cm$^{3}$/sec could create a small enhancement, it was not sufficient to push the line wings into emission. With energy deposition rates of 300 - 700 ergs/cm$^{3}$/sec, the H$\alpha$ line wing enhancement was 150\% that of the background profile, while a deposition rate of 1000 ergs/cm$^{3}$/sec appeared to more than double the line wing emission in H$\alpha$, while also enhancing the continuum by 20\%, which is contrary to observations.

An energy deposition rate of 500 ergs/cm$^{3}$/sec was thought to be sufficient to replicate an EB-like enhancement in the wings of H$\alpha$. This energy was then injected over 200~km, and placed across various regions of the photosphere/temperature minimum region in an attempt to locate the formation height of EBs (Grid \textbf{2}). We also modeled the emission in the Ca II 8542 A line, because, by considering two lines formed at different heights, we can better constrain the location of EB energy deposition. Fig. \ref{fig1} shows the response of the atmosphere to the injected energy, along with the corresponding line profiles.  

Since the density decreases with increasing height, depositing energy higher in the atmosphere results in greater temperature enhancements. A maximum temperature of 9000 K was generated in Grid \textbf{2}. Bi-directional flows are also formed, with peak velocities up to 5~km s$^{-1}$. These values are stronger than that of Grid \textbf{1}. The electron density increase is not linear with the height of injected energy. The increase appears to plateau around 1.5 x 10$^{13}$ cm$^{-3}$. This non-linear behaviour is also apparent in the corresponding H$\alpha$ and Ca II 8542~\AA\ line profiles, with the greatest difference in the wing emission being at the lower injected heights within Grid \textbf{2}. A small H$\alpha$ core enhancement is also apparent in the simulations. This effect could be greatly reduced with the application of a regular chromosphere \citep{robhan}. The lack of an overlying chromospheric canopy makes the small core contribution from the photosphere visible. 

\begin{figure*}[!t]
\centering
  \centering
  \includegraphics[width=.45\linewidth]{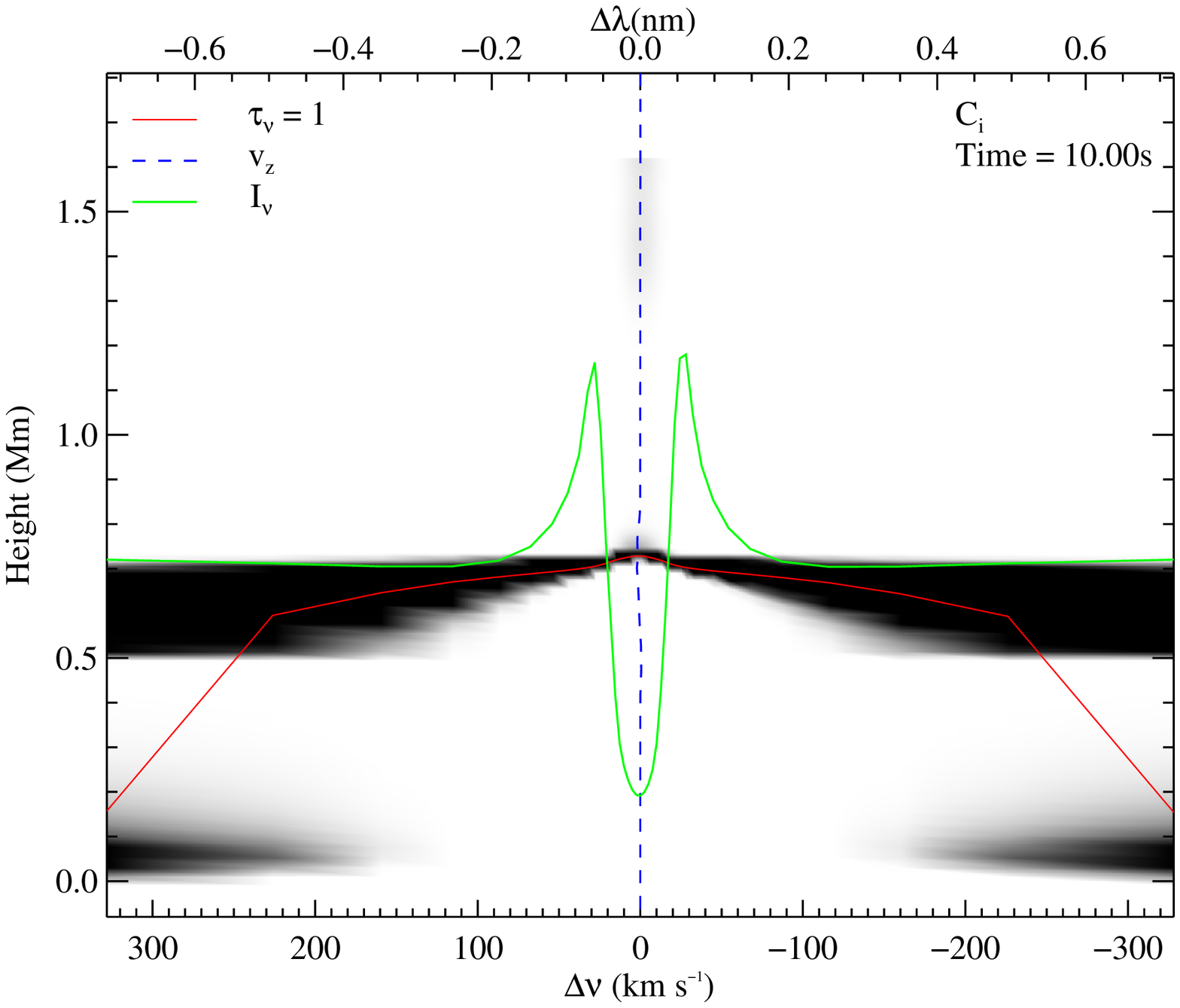}
  \label{fig:sub1}
  \centering
  \includegraphics[width=.45\linewidth]{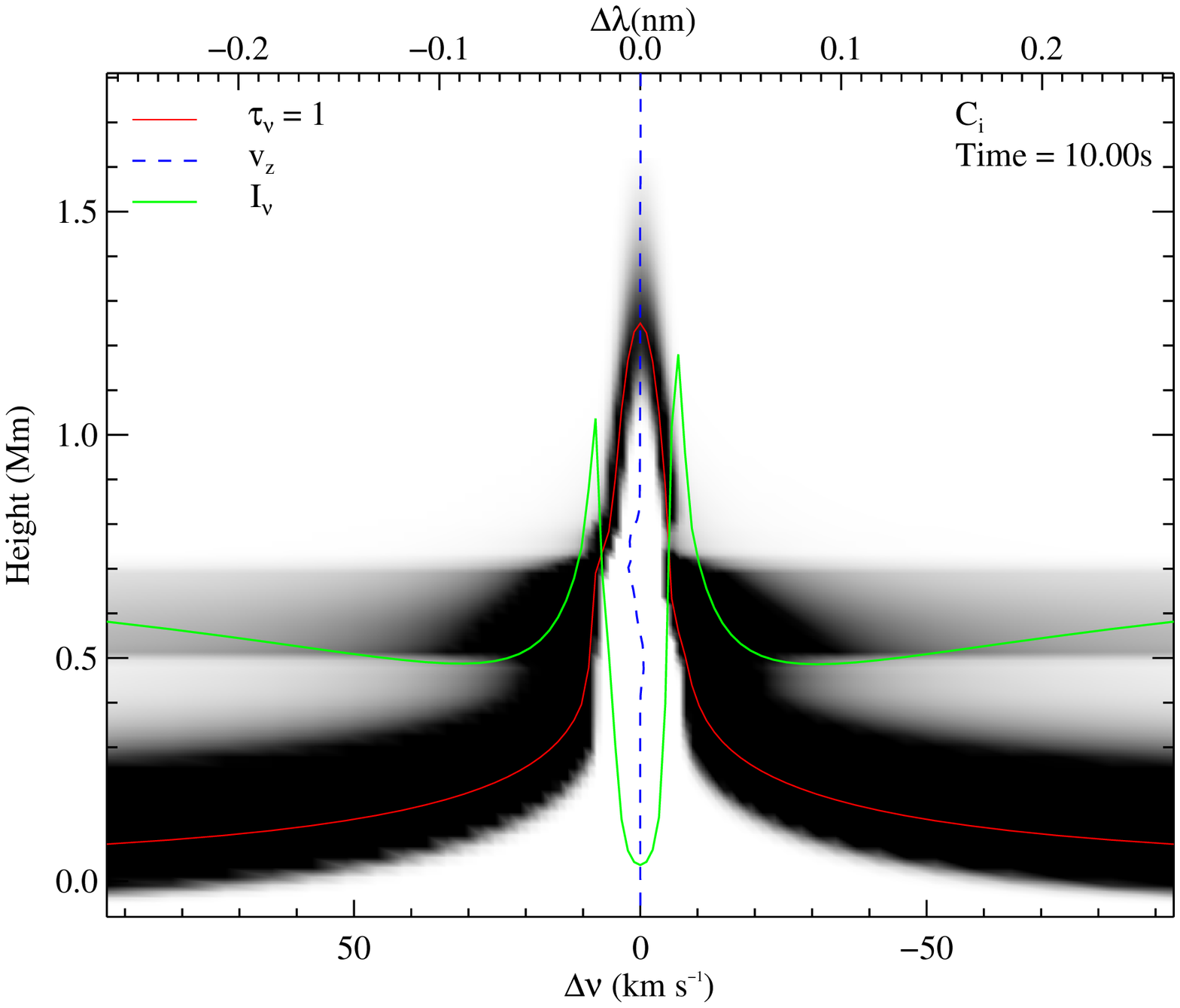}
  \label{fig:sub2}
\caption{Left: Contribution function for the H$\alpha$ line. Right: Contribution function for the Ca II 8542~\AA\ line profile. Dark regions indicate areas of strong contribution. The green lines show the line profiles. The red lines show the $\tau$ = 1 location, and the blue dashed line shows the vertical velocity as a function of height. }
\label{fig2}
\end{figure*}

Fig. \ref{fig2} shows contribution functions for the H$\alpha$ and Ca II 8542~\AA\ lines when an energy of 500 ergs/cm$^{3}$/sec is deposited between 500 - 700 km above the photospheric floor. When the energy is deposited higher in the atmosphere, a strong Ca II 8542~\AA\ core enhancement is apparent (see Fig. \ref{fig1}). When the energy is injected lower in the atmosphere, the wing enhancement broadens and becomes less intense overall, as the energy is being injected into a region where the outer wings are formed over a wider range of heights.

\section{Mg II h \& k line modeling}

We have found that energy deposited in the 500 - 700 km layer resulted in the best match to typically observed EB line profiles for the H$\alpha$ and Ca II 8542~\AA\ lines (e.g. Fig. 5 of \cite{viss}). In this Section, we consider the response of the Mg II h \& k lines to similar EB heating. Recent observations with the IRIS instrument hint at enhancements in the wings of these lines (see Fig. 10 of \citep{viss2}). We use the RH line synthesis code \citep{uit, tiago} to calculate the line profiles in partial redistribution (PRD). The 2 grids of models run for H$\alpha$ and Ca II in the previous section produce no enhancement in the Mg II h \& k lines (Grids \textbf{1} and \textbf{2}).

A further grid of models was created, which increases the energy deposited around the temperature minimum exponentially, between 1000 - 1 x 10$^{6}$ ergs/cm$^{3}$/sec (Grid \textbf{3}). Again, no enhancement is observed in the Mg II h \& k lines, though strong continuum enhancement appears with these high energy deposition rates. The atmospheric response to this sudden energetic injection did not create dramatically higher temperatures. In fact, the peak temperature in the 1 x 10$^{6}$ ergs/cm$^{3}$ model reached 9500~K, with a corresponding electron density of 10$^{14}$ cm$^{-3}$ and a peak shock velocity of 5 km s$^{-1}$. 

The contribution function of the Mg II h \& k lines appeared to show a line core formation height concentrated in the upper chromosphere, with little contribution from the photosphere. The wings of the lines were formed slightly lower, closer to the line core formation region of Ca II 8542~\AA. If the energy deposition occurs in the lower-mid chromosphere, enhanced emission in the Mg II h \& k wings can be achieved, though this will only cause flare-like profiles producing full emission in Ca II 8542~\AA\ and H$\alpha$ (Grid \textbf{4}). 

It also became apparent that the background Mg II h \& k line profile of our model appeared in absorption rather than in emission as is observed. In an attempt to rectify this issue, a different starting atmosphere was adopted which has it's transition region pushed to higher column masses, to reflect a more active, plage-like atmosphere (QS.SL.HT from \citep{allred}). 

With this new starting atmosphere, the initial profile of the Mg II h \& k lines appear in emission, similar to the background profiles observed \citep{Peter, viss2, kim, tian}. However, the line cores of Ca II 8542~\AA\ and H$\alpha$ are also slightly enhanced, with Ca II 8542~\AA\ being more prominent. 

Again, two grids of models were run, one grid applying 1000 ergs/cm$^{3}$/sec of heating in 200~km depths over the photosphere and temperature minimum, similar to Fig. \ref{fig1} (Grid \textbf{5}). The second grid applied the same heating function in the chromosphere, up to 2~Mm above the photospheric floor (Grid \textbf{6}). We find that injecting the energy into the chromosphere now only causes strong line core emission in the H$\alpha$, Ca II 8542~\AA, and Mg II h \& k line profiles. This is accompanied by large Doppler shifts due to very strong velocities ($>$20 km s$^{-1}$) in the leading shock front. The strong Doppler velocities appear due to the shock front interacting with transition region which is now at higher column masses. This forces the transition region to shift even lower in the atmosphere.

\begin{figure*}[!t]
\plotone{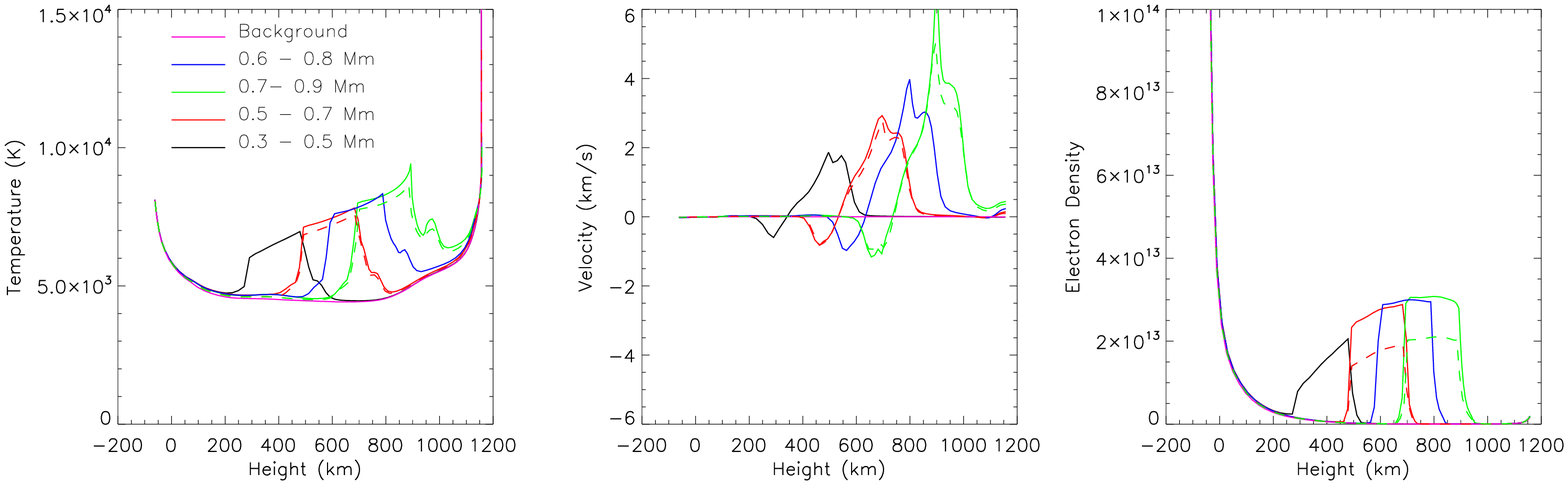}
\caption{Left: Temperature profile across the temperature minimum region for 1000 ergs/cm$^{3}$/sec deposited at various heights, using the pre-flare starting atmosphere. Middle: Vertical velocity profiles. Right: Electron density profiles. The dashed lines indicate runs with the injected energy density halved.}
\label{fig3}
\end{figure*}

In Grid \textbf{5}, in which we varied the energy deposition location over the photosphere and temperature minimum region, we find a similar atmospheric response to that of Fig. \ref{fig1} (Grid \textbf{2}), only slightly stronger due to the doubled heating rate. This can be seen in Fig. \ref{fig3}.

\begin{figure*}[!t]
\plotone{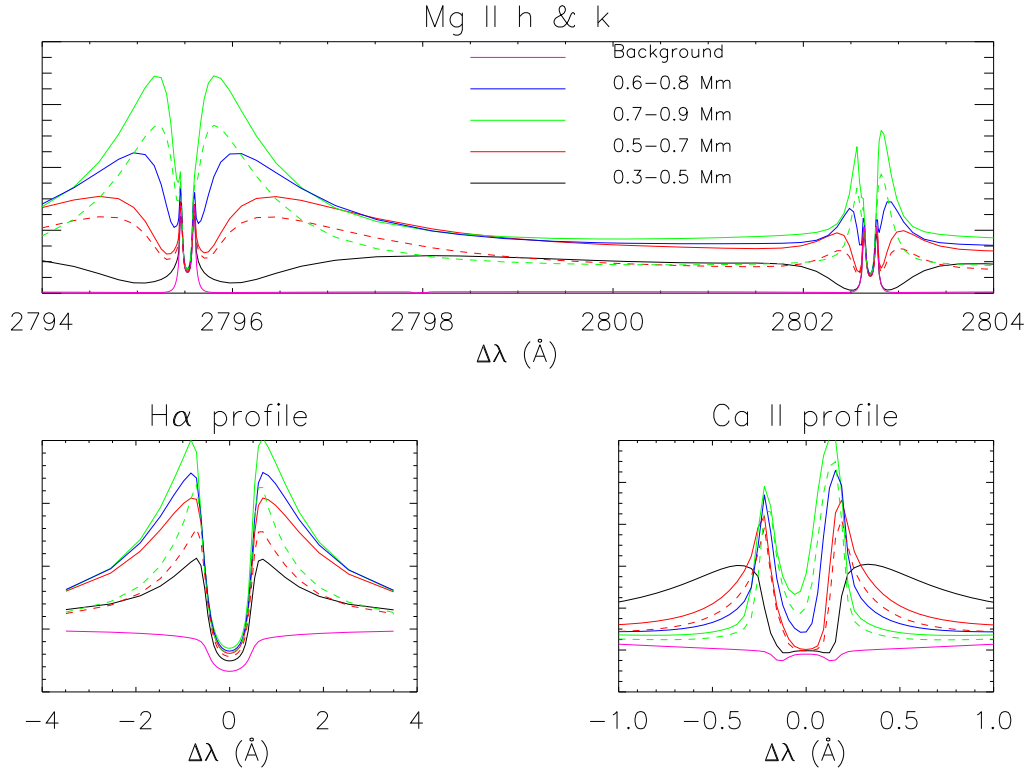}
\caption{Top: Mg II h \& k line profiles, calculated in PRD from the atmospheres in Fig. \ref{fig3}. Bottom Left: H$\alpha$ line profiles. Bottom Right: Ca II 8542~\AA\ line profiles. The dashed lines indicate the line profiles when the energy deposition is halved.}
\label{fig4}
\end{figure*}

Fig. \ref{fig4} shows the corresponding line profiles synthesised for the atmospheres shown in Fig. \ref{fig3}.  When the energy is deposited below the temperature minimum region (300 - 500 km above the photospheric floor), there is very little response in all synthesised lines, but a noticeable continuum increase, and slight wing enhancement in H$\alpha$ and Ca II 8542~\AA. Similar to the quiet Sun model, the best fitting for the H$\alpha$ and Ca II 8542~\AA\ line profiles appears when the energy is inserted around the temperature minimum region (500 - 700 km). However, the Mg II h \& k lines appear to have only their outer wings showing enhancement. The response of Mg II h \& k when the heating is applied between 700 - 900 km above the photospheric floor is most similar to that of strong observed IRIS bombs (see Fig. 10 in \citep{viss2}). However, the Ca II 8542~\AA\ line profile shows strong core emission with energy deposited at this height. When halving the energy deposition rate to the ideal rate identified in the previous section, we find the enhancements in the lines reduce slightly in Ca II 8542~\AA\  and Mg II h \& k, with a slightly larger differential apparent in the H$\alpha$ line profile.

To investigate if a larger depth range for the energy deposition is more appropriate, we also ran two larger models in both the quiet Sun and pre-flare atmospheres (Grids \textbf{7} and \textbf{8}). These models inject 500 ergs/cm$^{3}$/sec over ranges of 500 - 1000 km and 500 - 1200 km. In all cases, strong core enhancements were seen in the synthesised Ca II 8542~\AA\ and H$\alpha$ line profiles. The Mg II h \& k line profiles appeared to show flare-like profiles with the pre-flare atmosphere, and strong enhancement in the line wings in the quiet Sun atmosphere.

\section{Discussion and Conclusions}

We have used 1-dimensional radiative hydrodynamical simulations of the solar atmosphere to investigate the location and rate of energy deposition required to reproduce EB-like line profiles. Our findings based on a quiet Sun starting atmosphere suggest that the  location of energy deposition is near the temperature minimum region, around 600~km above the photospheric floor. We find that placing the energy into the chromosphere causes strong emission, especially in the core of Ca II 8542~\AA. When considering the Mg II h \& k line profiles, the quiet Sun starting atmosphere is not viable, as it produced a background profile in absorption. Inserting the energy around the temperature minimum only enhanced the continuum around these lines. The mid-upper chromosphere was the location of the main contribution of the Mg II h \& k lines, and so only by depositing energy here could one get a visible response in these lines, which could overcome the continuum enhancement. This would then only cause flare-like profiles in the H$\alpha$ and Ca II 8542~\AA\ line cores.

 A pre-flare, plage-like starting model was used in an attempt to overcome this and push the formation to lower atmospheric heights. While this was the case, the H$\alpha$ and Ca II 8542~\AA\ line cores were enhanced in comparison to the quiet Sun model, with the Ca II 8542~\AA\ background profile appearing particularly unrealistic in comparison to observations.

However, it was possible to achieve Mg II h \& k line profiles similar to observations by depositing 500 ergs/cm$^{3}$/sec at 700-900 km above the photospheric floor. While the Mg II profiles look similar to IRIS observations, the Ca II 8542~\AA\ cores appear in emission.  If the energy is inserted 200 km lower in the atmosphere (the ideal setup from the quiet Sun models), the H$\alpha$ and Ca II 8542~\AA\ line profiles appear EB-like, with wing emission and no core enhancement, relative to the background profile. The Mg II h \& k profiles however appear with enhancement in the outer wings, much unlike observations. 

We have shown that by considering the starting atmosphere to be quiet Sun, we can obtain EB-like synthetic H$\alpha$ and Ca II 8542~\AA\ line profiles. This results in bi-directional flows at the EB location due to the formation of a shock, with a stronger up-flow (2 - 5 km s$^{-1}$), and weaker trailing down-flow (1 - 2 km s$^{-1}$). This fits with observations reporting bi-directional jets at EB locations \citep{matsu2, wat1} and previous EB modeling attempts \citep{ach}. A local temperature enhancement of 2000 - 3000 K also forms, and fits with previous estimates \citep{geo,fang,iso,ach, li}. This situation however did not accurately model the Mg II h \& k lines, which needed a more active starting atmosphere. While we could recreate Mg II h \& k profiles similar to recent IRIS observations, we were unable to attain EB-like line profiles in all 3 of our diagnostic lines simultaneously. We speculate that an EB-specific atmospheric model may be needed, which has different atmospheric abundances, enabling the Ca II 8542~\AA\ line core and Mg II h \& k line wings to form at different heights. We also find that we were unable to super-heat the plasma to $>$10,000~K as previous works speculate in order to explains the Mg II h \& k line formation \citep{Peter, viss2, kim, tian}.

AR thanks M. Carlsson for valuable comments on the paper. Research at the Armagh Observatory is grant-aided by the Northern Ireland Department of Communities. We acknowledge support by STFC. This research was supported by the SOLARNET project (www.solarnet-east.eu), funded by the European Commissions FP7 Capacities Program under the Grant Agreement 312495. The research leading to these results has received funding from the European Communitys Seventh Framework Programme (FP7/2007-2013) under grant agreement no. 606862 (F-CHROMA).

\begin{table*}[!t]
\centering
\begin{tabular}{| c | c | c | c | c |}
\hline
 \textbf{Grid Number} & \textbf{Energy Rate} & \textbf{Height (km)} & \textbf{Atmosphere} & \textbf{Comments} \\ \hline
1 & 100 - 1000 & 300 - 500 & Quiet Sun & 300 - 700 ergs/cm$^{3}$/sec best fit EB profiles in H$\alpha$.  \\ \hline
2 & 500 & 300 - 1000 (200~km step) & Quiet Sun & 500 - 700 km best fit EB profiles in H$\alpha$ and Ca II. \\ \hline
3 & 1000 - 1 x 10$^{6}$& 300 - 500 & Quiet Sun & All $<$10,000~K. Overwhelmed by continuum emission in all lines.\\ \hline
4 & 500 & 1000 - 2000 (200~km step) & Quiet Sun & Flare-like profiles in Ca \& H$\alpha$. Good Mg II response. \\ \hline
5 & 500/1000 & 300 - 900 (200~km step) & Plage & 500 - 700 km best fits EB Ca \& H$\alpha$ profiles. 700 - 900 km best fits IB Mg II profiles. \\ \hline
6 & 500/1000 & 1000 - 2000 (200~km step) & Plage & Flare-like profiles in all lines. \\ \hline
7 & 500 & 500 - 1000 & Both & Strong line core enhancement in Ca, H$\alpha$. Weak Mg II response in QS atmosphere.  \\ \hline
8 & 500 & 500 - 1200 & Both & Ca in full emission, H$\alpha$ strong enhancement. Good Mg II response in QS atmosphere. \\ \hline
\end{tabular}
\caption{The grids of models used in this study. IB = IRIS bombs. QS = Quiet Sun starting atmosphere. Energy rate is in ergs/cm$^{3}$/sec. }
\label{table1}
\end{table*}

\end{document}